%% file: WPCF_pol.tex
\documentclass[a4paper,10pt]{article}

\usepackage[T1]{fontenc}
\usepackage[utf8]{inputenc}
\usepackage[unicode]{hyperref}
\usepackage[table,xcdraw]{xcolor}
\usepackage[font=small,labelfont=bf]{caption}
\usepackage{array,float,multirow,hhline,booktabs,amsmath,amssymb,gensymb,anysize,geometry,mathtools,subcaption,authblk}
\usepackage[pdftex]{graphicx}
\usepackage[bold]{hhtensor}
\usepackage[noadjust]{cite} 

\title{Analytic model studies of polarized baryon production}
\author[1]{B\'alint~Boldizs\'ar}
\affil[1]{Department of Atomic Physics, E{\"o}tv{\"o}s Lor{\'a}nd University,  P{\'a}zm{\'a}ny P. s. 1/A, H-1117 Budapest, Hungary;
bolbalaa@caesar.elte.hu}
\date{}
\include{mycommands}

\begin{document}
\maketitle

\begin{abstract}
We investigate known exact solutions of hydrodynamics and derive analytic formulas for the polarization of baryons produced at freeze-out. Such polarization
(observed in high energy heavy-ion experiments) carries information on the time evolution of the quark-gluon plasma (sQGP), and our results give first
analytic insight into the connection between this type of measurements and dynamical properties of the sQGP (e.g. vorticity). We present results for a
rotating and acceleratingly expanding solution and also give hints on how to calculate the polarization using a rotating extension of the Buda--Lund
parameterization.
\end{abstract}

\section{Introduction}

Observation of net polarization of $\Lambda$ baryons in high energy heavy-ion collisions by the STAR experiment~\cite{STAR:2017ckg} fall in line with
expectations predicting such polarization if local thermal equilibrium is assumed also for spin degrees of freedom~\cite{Becattini:2013fla}. The main
interest in this topic is because by measuring such polarization an insight is gained into the details of the expansion dynamics of the strongly
interacting Quark-Gluon Plasma (sQGP). As an example, non-vanishing polarization may be a consequence of rotating expansion of the sQGP, and the
time evolution of this rotation is in connection to the Equation of State (EoS) of the sQPG.

Numerical model studies of polarization~\cite{Csernai:2014nva,Xie:2016fjj,Karpenko:2016jyx,Xie:2017upb} indeed predict non-zero polarization
and connect this quantity to properties (e.g. vorticity) of the flow. However, numerical simulations by their nature do not always give a clear
picture of the dependence of final state observables on assumptions made on the initial state.

In the work presented here~\cite{Boldizsar:2018akg} we set out to give analytic formulas for the polarization of massive spin 1/2 particles, based on the
simple formula written up in Ref.~\cite{Becattini:2013fla} under the assumption of local thermal equilibrium. This type of investigations can lead to a
straightforward connection between properties of the flow and experimentally observable quantities (in our case polarization). We utilize known exact
analytic solutions of perfect fluid hydrodynamics that grab some aspects of the real geometry of a heavy-ion collision. In the following we briefly present
the calculations leading to results about polarization and illustrate the main findings. Additional details are to be found in Ref.~\cite{Boldizsar:2018akg}.

\section{Basic equations and assumptions}

In phenomenological modelling of final state observables, the source function $f(x,p)$ that describes the distribution of particles produced in
the hadronization process can be calculated from a thermal ensemble that corresponds to the final state of the hydrodynamical evolution of the sQGP.
For spin 1/2 particles, $f(x,p)$ is a locally thermal Fermi--Dirac distribution. In Ref.~\cite{Becattini:2013fla} the following formula is established
for the spin vector of locally thermally equilibrated fermions: 
\begin{align}
\label{e:polarization}
\langle S(p) \rangle^\mu &= \frac{\int \m d^3 \Sigma_\nu p^\nu f(x,p) \langle S(x,p) \rangle^\mu}
{\int \m d^3\Sigma_\nu p^\nu f(x,p)},&
\langle S(x,p) \rangle^\mu &= \frac{1}{8m}\big(1 {-} f(x,p)\big) \varepsilon^{\mu\nu\rho\sigma} p_\sigma \partial_\nu \beta_\rho.
\end{align}
Here $\langle S(x,p) \rangle^\mu$ is the space-time and momentum dependent spin vector of the produced particles, which is averaged over the freeze-out
with the $f(x,p)$ distribution to get the observable (momentum dependent) polarization $\langle S(p) \rangle^\mu$. Other notations are:
$p^\mu{\equiv}(E,\v p)$ and $m$ are the momentum and the mass of the particle (with $E^2 {=} m^2{+}\v p^2$ and $c{=}1$). $T(x)$, $u^\mu(x)$, and $\mu(x)$ are
the temperature, four-velocity, and chemical potential fields of the fluid, respectively. We also customarily introduced the inverse
temperature field as $\beta^\mu \equiv u^\mu /T$.

Numerical calculations evaluate these formulas while solving the equations of hydrodynamics at the same
time. On the other hand, we can take known exact solutions and directly evaluate the formula for $\langle S(p) \rangle^\mu$ given above. As a proof of
concept, we first investigate the case of spherically symmetric Hubble flow described in Ref.~\cite{Csorgo:2003ry}. We calculate $\obs{S(p)}^\mu$ in an exact
accelerating and rotating relativistic solution~\cite{Nagy:2009eq,Hatta:2014gqa}). As an outlook we take a glance at the Buda--Lund model (see e.g.
Ref.~\cite{Csanad:2003qa}): we write up a rotating generalization and specify some formulas for the polarization in this model case. 
 
We evaluate the polarization from Eq.~\eqref{e:polarization} for a given $\beta^\mu$ field, which is specified by a particular hydrodynamical solution.
In the expression for $S(x,p)$, Eq.~\eqref{e:polarization}, we made the $f\ll 1$ assumption (corresponding to the Maxwell-Boltzmann limiting case,
which is usually assumed for calculations of final state observables). Also, in our calculations we used the saddle-point integration method;
with this approximation the space-time averaged observable polarization vector becomes simply  
\begin{align}
\label{e:polsimple}
\langle S(x,p) \rangle^\mu = \rec{8m}\varepsilon^{\mu\nu\rho\sigma} p_\sigma \partial_\nu \beta_\rho
\Follows \langle S(p) \rangle^\mu \approx \rec{8m}\varepsilon^{\mu\nu\rho\sigma} p_\sigma \partial_\nu \beta_\rho \Big|_{\v r=\v R_0},
\end{align}
where $\v R_0$ is the position of the saddle point (the point of maximum emittivity), which depends on $\v p$. So to get analytic formulas for the
polarization for a given $\beta^\mu(x)$ field and freeze-out condition, one has to express $\v R_0$ as a function of particle momentum $\v p$ and evaluate
the $\obs{S(x,p)}^\mu$ at this position. As usual in heavy-ion phenomenology, we also neglect the $\mu/T$ (fugacity) factor in the expression of the source
function:
\begin{align}
\label{e:MB1}
f(x^\mu ,p^\mu) = \frac{g}{(2\pi\hbar)^d}\exp\big({-}p_\mu \beta^\mu\big),\qquad\textnormal{where again:}\quad \beta^\mu = \frac{u^\mu}{T}.
\end{align}
Below we consider two exact solutions of relativistic perfect fluid hydrodynamics and outline the calculation of the polarization
of produced spin 1/2 baryons in these models, using the methods and approximations described above. For a more detailed discussion see
Ref.~\cite{Boldizsar:2018akg}; here we summarize the main steps. 

\section{Polarization in exact analytic hydrodynamical solutions}

One of the solutions considered is a simple spherically symmetric special case of the (more general, ellipsoidal) Hubble-type solution (presented in its
fullest form in Ref.~\cite{Csorgo:2003ry}), which is characterized by the following velocity ($u^\mu$), temperature ($T$) and density ($n$) profiles:
\begin{align}
\label{e:hubble}
u^\mu & = \frac{x^\mu}{\tau},&
n &= n_0 \left(\frac{\tau_0}{\tau}\right)^d,&
T &= T_0 \left(\frac{\tau_0}{\tau}\right)^{d/\kappa}.
\end{align}
Here $d=3$ is the dimensionality of space, and as usual, we use $\tau = \sqrt{t^2-\v r^2}$, and $\kappa = 1/c_s^2$ is the inverse square of the speed of
sound, assumed to be constant (this constant appears in the Equation of State as $\varepsilon/p$).

The freeze-out is assumed to happen on a constant $\tau=\tau_0$ hypersurface. The Cooper-Frye prefactor for this assumption (at a given
point on the hypersurface whose spatial coordinate is $\v r$) turns out to be
\begin{align}
\label{e:mb}
t(\v r)\equiv\sqrt{\tau_0^2{+}\v r^2}\follows \m d^3\Sigma_\mu = \rec{t(\v r)}\begin{pmatrix} t(\v r) \\ \v r \end{pmatrix}\m d^3\v r.
\end{align}
In the case of the spherically symmetric Hubble flow, the temperature takes a constant value on the freeze-out hypersurface (denoted here by $T_0$), and
the position of the point of maximum emittivity $\v R_0$ as well as the $\partial_\nu\beta_\rho$ derivative (a necessary ingredient in
Eq.~\eqref{e:polsimple} for the calculation of the polarization) can be calculated as
\begin{align}
p^\mu\beta_\mu = \frac{ E t(\v r){-}\v p{\v r}}{T_0} \follows R_0=\frac{\tau_0}{m}\v p, \qquad\quad
\partial_\nu \beta_\rho = \frac{g_{\nu\rho}}{\sqrt{\tau_0^2 {+} r^2}T_0}+\frac{r_\nu r_\rho}{(\tau_0^2 {+} r^2)^{3/2}T_0}.
\end{align}
With some simplifications, one can verify that both the time and spatial components of the resulting polarizaton four-vector are zero:
\begin{align}
&\langle S(p) \rangle^0 =\frac{1}{8mT_0} \varepsilon^{0ikl} p_l\partial_i \beta_k\bigg|_{\v r =\v R_0} = \rec{8mT_0} \varepsilon_{ikl}p_l
\bigg(\frac{g_{ik}}{\sqrt{\tau_0^2 {+} r^2}T_0}+\frac{r_{i}r_{k}}{(\tau_0^2 {+} r^2)^{3/2}T_0}\bigg)\bigg|_{\mathbf r {\,=\,} \mathbf R_0}=\ldots=0,\\
&\langle S(p)\rangle^i = \rec{8mT_0}
\bigg(\varepsilon_{ikl}p_l\partial_k\beta_0-\varepsilon_{ikl}p_l\partial_0\beta_k-\varepsilon_{ikl}p_0\partial_k\beta_l\bigg)\bigg|_{\v r=\v R_0}=\ldots=0.
\end{align}
In conclusion, the polarization four-vector in the spherical symmetric self-similar flow is
\begin{align}
\langle S(p)\rangle^\mu = \begin{pmatrix} 0\\ \mathbf 0\\ \end{pmatrix},
\end{align}
which is consistent with our expectations, since this solution is totally spherically symmetric.

A more realistic (and for our goals, more interesting) solution to be studied is a rotating and accelerating expanding solution, first written up
in Ref.~\cite{Nagy:2009eq}: 
\begin{align}
\label{e:rot}
\v v &= \frac{2t\v r {+} \tau_0^2\gvec\Omega{\times}\v r}{t^2{+}r^2{+}\rho_0^2},&
T&= \frac{T_0\tau_0^2}{\sqrt{(t^2{-}r^2{+}\rho_0^2)^2{+}4\rho_0^2 r^2{-}\tau_0^4(\gvec\Omega{\times}\v r})^2},&
n &= n_0\z{\frac{T}{T_0}}^3.
\end{align}
Here the $\rho_0$ parameter characterizes the initial spatial extent of the system. The $T_0$ and $\tau_0$ (the freeze-out values) are included for the
sake of consistency of physical units, and $\gvec\Omega$ is an angular velocity three-vector indicating the axis and magnitude of rotation.
We write up the $\beta^\mu$ field as follows:
\begin{align}
\label{e:solclass}
&\beta^\mu = \frac{u^\mu}{T} = a^\mu {+} F^{\mu\nu}x_\nu {+} (x^\nu b_\nu) x^\mu {-} \frac{x^\nu x_\nu}{2}b^\mu,
\\\textnormal{with}\quad&
a^\mu {\,=\,} \frac{\rho_0^2}{2T_0\tau_0^2}\begin{pmatrix}1\\\mathbf 0\end{pmatrix},\qquad b^\mu {\,=\,} \rec{T_0\tau_0^2}\begin{pmatrix}1\\\mathbf 0\end{pmatrix}, \qquad
F_{0k} {\,=\,} F_{k0} {\,=\,} F_{00} {\,=\,} 0, \qquad F_{kl} {\,=\,} \varepsilon_{klm} \frac{\Omega_m}{2T_0}.
\end{align}
We need to find the saddle point $\v R_0$; the result is:
\begin{align}
&p_\mu\beta^\mu = \rec{T_0\tau_0^2}\Big(E(2r^2{+}\tau_0^2{+}\rho_0^2)-2\sqrt{\tau_0^2{+}r^2}\v r\v p-\tau_0^2\v r(\v p{\times}\gvec\Omega)\Big),\qquad
\nabla\Big\{p_\mu\beta^\mu\Big\}\Big|_{\v r = \v R_0} \stackrel{!}{\,=\,}0\followse\nonumber\\
&\bfollows \v R_0 = \frac{\tau_0}{2p}\sqrt{\frac{E{-}m}{2m}}\sqrt{\tau_0^2(\hat{\v p}{\times}\gvec\Omega)^2(E{-}m)^2+4p^2}\cdot\hat{\v p} +
\tau_0^2\frac{E{-}m}{2p}\cdot\hat{\v p}{\times}\gvec\Omega,\qquad\textnormal{with}\quad \hat{\v p}:=\frac{\v p}{|\v p|}.
\label{e:R0eq:rotaccel}
\end{align}
The $\beta_\mu$ field from Eq.~\eqref{e:solclass} is thus used for the calculation of the polarization following Eq.~\eqref{e:polsimple}
as
\begin{align}
\partial_\nu \beta_\rho =
  F_{\rho\nu} {+} x^\alpha b_\alpha g_{\nu\rho} {+} x_\rho b_\nu {-} x_\nu b_\rho \quad\follows\quad
\langle S(p)\rangle^\mu =
  \rec{8m}\varepsilon^{\mu\nu\rho\sigma}p_\sigma
  \Big(F_{\rho\nu} {+} x_\rho b_\nu {-} x_\nu b_\rho \Big)\Big|_{\mathbf r {\,=\,} \mathbf R_0}.
\end{align}
Evaluating this by substituting the expressions of $F_{\mu\nu}$ and $b_\mu$, and collecting the time-like and space-like components carefully,
we finally get the following concise result for the polarization four-vector in the case of the rotating and accelerating solution:
\begin{align}
\label{e:forgopol}
\langle S(p)\rangle^\mu =\frac{1}{8mT_0} \begin{pmatrix} \mathbf p\gvec\Omega \\ m\gvec\Omega +\frac{E{-}m}{p^2}(\gvec\Omega\mathbf p)\mathbf p \end{pmatrix}.
\end{align}
In case of $\gvec\Omega = 0$, there is no rotation. Indeed we get $\langle S(p)\rangle^\mu{=}0$ in this case: in this model, polarization is very
transparently connected to the presence of rotation.

\begin{figure}[H]
\centering
\includegraphics[width=\textwidth]{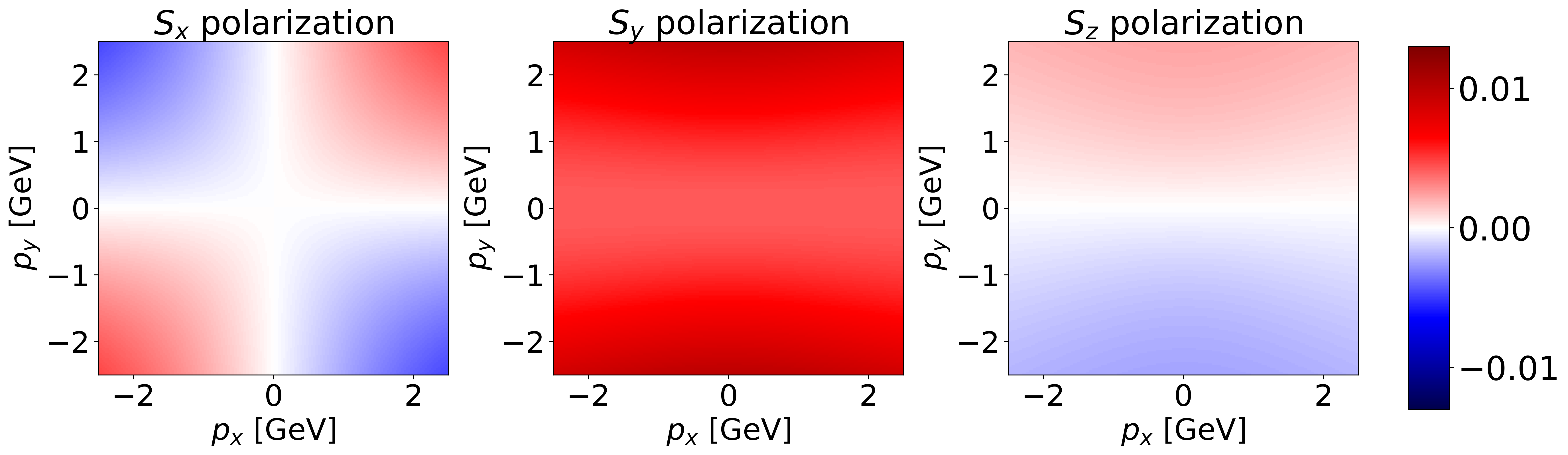}
\caption{Components of the polarization vector for the rotating and accelerating solution according to Eq.~\eqref{e:forgopol}.
The mass $m$ is taken as $m {=} m_\Lambda{=}1115$~MeV$/c^2$, and a realistic $|\gvec\Omega| = 0.1\,c/\m{fm}$ value was chosen for this illustration.}
\label{f:pol}
\end{figure}

Transforming the polarization vector into the rest frame (``r.f.'') of the particle, the result is
\begin{align}\label{e:polSrf}
\langle S(p) \rangle^\mu_{\textnormal{r.f.}} = \begin{pmatrix} 0 \\ \mathbf S_{\textnormal{r.f.}} \end{pmatrix},
\quad\textnormal{where}\quad \mathbf S_\textnormal{r.f.} = \rec{8T_0}\gvec\Omega. 
\end{align}
In this case it thus turns out that the polarization in the rest frame of the particle is independent of momentum. The helicity of the produced particles
in this case is (the $\mathbf S$ polarization vector is taken in the laboratory frame)
\begin{align}
\label{e:helicity}
H := \hat{\v p}\v S = \frac{E}{8mT_0}\hat{\v p}\gvec\Omega.
\end{align}
Fig.~\ref{f:pol} illustrates these results (in the same way as it became somewhat customary for numerical simulations): we plot specific components of the
polarization vector in the laboratory frame as a function of momentum components, in the $p_z=0$ transverse plane.
\begin{center} * * * \end{center}
Having seen that it is indeed possible to obtain simple analytic formulas for the polarization in simple exact hydrodynamical solutions (although not fully
realistic ones), we turn our attention to a logical next step of this line of investigation. We consider the Buda--Lund model~\cite{Csorgo:1995bi,
Csanad:2003qa}: a more involved hydrodynamical final state parametrization that is successful in describing usual one-particle and two-particle observables
(soft spectra and correlations). We present some first formulas on how to calculate the polarization in this model; this investigation can also
unveil the dependence of the polarization on other factors beyond rotation (such as temperature gradient and/or the acceleration of the expansion).

We write up a simple rotating generalization of the ellipsoidal Buda--Lund model as
\begin{align}
u^\mu = \begin{pmatrix} \sqrt{1+\v u^2} \\ \v u \end{pmatrix},\qquad\textnormal{where}\quad
\v u = \begin{pmatrix}
\frac{\dot X(t)}{X(t)} r_x+ \omega(t)\frac{R(t)}{Z(t)}r_z\\
\frac{\dot Y(t)}{Y(t)} r_y\\
\frac{\dot Z(t)}{Z(t)} r_z -\omega(t)\frac{R(t)}{X(t)} r_x
\end{pmatrix},\qquad \textnormal{and } \quad R(t)\equiv \frac{X(t){+}Z(t)}{2}.
\end{align}
Here $\omega$ is the newly introduced angular velocity, and the $X$, $Y$, $Z$ and $\dot X$, $\dot Y$, $\dot Z$ parameters correspond to
the principal axes and expansion velocity components of the ellipsoid-like source. The temperature field is
\begin{align}
\rec{T} = \frac{1{+}a^2A}{T_0}\z{1+\alpha^2\frac{(\tau{-}\tau_0)^2}{2(\Delta\tau)^2}},\qquad\textnormal{with}\quad
a^2\equiv \frac{T_0{-}T_s}{T_s}, \quad \alpha^2\equiv\frac{T_0{-}T_e}{T_e},\quad A = \frac{r_x^2}{2X^2}{+}\frac{r_y^2}{2Y^2}{+}\frac{r_z^2}{2Z^2}.
\end{align}
The $\tau_0$ proper-time value corresponds to the freeze-out hypersurface (with ,,width'' $\Delta\tau_0$). The temperature values $T_0$, $T_s$ and $T_e$
are the values taken at the center of the ellipsoid, on the ,,surface'' of it, and the value after freeze-out, respectively. In this way, the $a^2$
parameter controls the temperature gradient. The $A$ function is called scaling variable; its constant values are coordinate-space ellipsoids.

\begin{figure}[H]
\centering
\includegraphics[width=\textwidth]{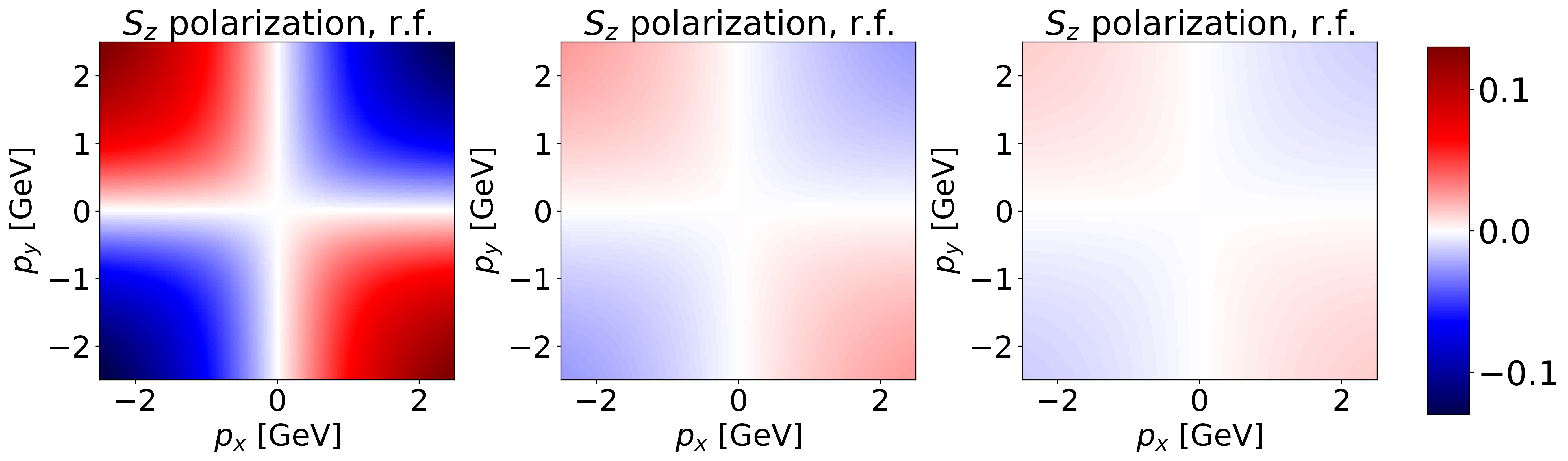}
\caption{$S_z$ component of the polarization vector from the rotating Buda--Lund model: with $a^2 = 0$ (left panel), $a^2 = 0.5$ (middle panel) and $a^2 = 1$
(right panel), meaning no temperature gradient, a moderate one, and a large temperature gradient, respectively. Other parameter values taken here are:
$m = m_\Lambda$, $\omega = 0.1\,c/\m{fm}$, $X=6$ fm, $Y=8$ fm, $Z=10$ fm, $\dot X = 0.3$, $\dot Y = 0.4$ and $\dot Z=0.2$.}
\label{f:blpol}
\end{figure}

The polarization can be calculated according to Eq.~\eqref{e:polsimple} as
\begin{align}
\obs{S(\v p)}^0 &=-\rec{8m}\varepsilon_{klm}\v p_m\partial_k\beta_l = -\rec{8m}\v p\z{\nabla\times\gvec\beta},\\
\obs{S(\v p)}^k &=\rec{8m}\Big((\partial_0\gvec\beta)\times\v p\Big)_k-\rec{8m}(\nabla\beta_0\times\v p)_k-\frac{p_0}{8m}(\nabla\times\gvec\beta)_k.
\end{align}
With a Lorentz boost we can express the polarization in the local rest frame of the produced particles:
\begin{align}
\v S_{\textnormal{r.f.}}&=\rec{8m}\kz{(\partial_0\gvec\beta-\nabla\beta_0)\times\v p-E(\nabla\times\gvec\beta)+\frac{E{-}m}{p^2}(\v p(\nabla\times\gvec\beta))\v p}.
\end{align}
In order to find the saddle point, one is led to a nonlinear system of algebraic equations; these can be solved either numerically or by successive
approximation. One then substitutes the resulting expression into the above formulas. Some very preliminary results from this investigation are shown
on Fig.~\ref{f:blpol}. We plot the $S_z$ component with respect to the momentum components (again for $p_z = 0$) for a given angular velocity $\omega=
0.1\,c/\m{fm}$, for three different $a^2$ values. 

\section{Summary}

We presented new analytic formulas for the polarization of baryons produced in high energy heavy-ion collisions, utilizing several analytic hydrodynamical
models. We considered a spherically symmetric self-similar flow (in which case the polarization is indeed zero, as is expected), and secondly,
a relativistic expanding and rotating hydrodynamical solution. In this latter case one gets simple formulas for the polarization that show a straightforward
connection of this quantity to the vorticity of the flow, however, the results and assumptions are too simple to be realistic. Finally, we presented
some preliminary investigations into the usability of the Buda--Lund model for calculating the polarization; from this much more complex model, one
can expect a realistic description of the experimentally observed polarization as well. We can thus reasonably hope that these studies have the potential
of a better understanding of polarization measurements and their phenomenological implications on the strongly coupled
Quark Gluon Plasma produced in heavy-ion collisions.

\section*{Acknowledgments}
The author is thankful to M\'arton~I.~Nagy and M\'at\'e~Csan\'ad for the numerous discussions and help given during the work presented here.
The author is grateful to the organizers of the 19th Workshop on Particle Correlations and Femtoscopy (WPCF 2019) conference.
This work was partially supported by the Hungarian NKIFH grants No. FK-123842 and FK-123959.

\end{document}

%% file: mycommands.tex
\ifdefined\myframe
\renewenvironment{myframe}[2]{\section{#1}\begin{frame}{#2}\vspace{-10pt}}{\end{frame}} 
\else
\newenvironment{myframe}[2]{\section{#1}\begin{frame}{#2}\vspace{-10pt}}{\end{frame}}
\fi

\ifdefined\mybullet
\renewcommand{\mybullet}[1]{\vspace{1mm}\\$\bullet$ {\bf #1}\vspace{1mm}\\}
\else
\newcommand{\mybullet}[1]{\vspace{1mm}\\$\bullet$ {\bf #1}\vspace{1mm}\\}
\fi

\ifdefined\mybulletEQ
\renewcommand{\mybulletEQ}[1]{$\bullet$ {\bf #1}\vspace{1mm}\\}
\else
\newcommand{\mybulletEQ}[1]{$\bullet$ {\bf #1}\vspace{1mm}\\}
\fi

\ifdefined\fracd
\renewcommand{\fracd}[2]{\frac{\displaystyle{#1}}{\displaystyle{#2}}}
\else
\newcommand{\fracd}[2]{\frac{\displaystyle{#1}}{\displaystyle{#2}}}
\fi

\ifdefined\recd
\renewcommand{\recd}[1]{\frac{\displaystyle 1}{\displaystyle{#1}}}
\else
\newcommand{\recd}[1]{\frac{\displaystyle 1}{\displaystyle{#1}}}
\fi

\ifdefined\pdd
\renewcommand{\pdd}[2]{\frac{\displaystyle{\partial{#1}}}{\displaystyle{\partial{#2}}}}
\else
\newcommand{\pdd}[2]{\frac{\displaystyle{\partial{#1}}}{\displaystyle{\partial{#2}}}}
\fi

\ifdefined\biggg
\renewcommand{\biggg}[1]{\scalebox{1.2}{\Bigg{#1}}}
\else
\newcommand{\biggg}[1]{\scalebox{1.2}{\Bigg{#1}}}
\fi

\ifdefined\Biggg
\renewcommand{\Biggg}[1]{\scalebox{1.4}{\Bigg{#1}}}
\else
\newcommand{\Biggg}[1]{\scalebox{1.4}{\Bigg{#1}}}
\fi

\ifdefined\eq
\renewcommand{\eq}[1]{\begin{equation}{#1}\end{equation}}
\else
\newcommand{\eq}[1]{\begin{equation}{#1}\end{equation}}
\fi

\ifdefined\eql
\renewcommand{\eql}[2]{\begin{equation}\label{e:#1}{#2}{\end{equation}}
\else
\newcommand{\eql}[2]{\begin{equation}\label{e:#1}{#2}\end{equation}}
\fi

\ifdefined\ali
\renewcommand{\ali}[1]{\begin{align}{#1}\end{align}}
\else
\newcommand{\ali}[1]{\begin{align}{#1}\end{align}}
\fi

\ifdefined\nl
\renewcommand{\nl}{{\vspace{0.2 cm}\\}}
\else
\newcommand{\nl}{{\vspace{0.2 cm}\\}}
\fi

\ifdefined\Tr
\renewcommand{\Tr}{{\mathrm Tr}}
\else
\newcommand{\Tr}{{\mathrm Tr}}
\fi

\ifdefined\BR
\renewcommand{\BR}{{\mathbb R}}
\else
\newcommand{\BR}{{\mathbb R}}
\fi

\ifdefined\BZ
\renewcommand{\BZ}{{\mathbb Z}}
\else
\newcommand{\BZ}{{\mathbb Z}}
\fi

\ifdefined\BN
\renewcommand{\BN}{{\mathbb N}}
\else
\newcommand{\BN}{{\mathbb N}}
\fi

\ifdefined\BC
\renewcommand{\BC}{{\mathbb C}}
\else
\newcommand{\BC}{{\mathbb C}}
\fi

\ifdefined\Re
\renewcommand{\Re}{\operatorname{Re}}
\else
\newcommand{\Re}{\operatorname{Re}}
\fi

\ifdefined\Im
\renewcommand{\Im}{\operatorname{Im}}
\else
\newcommand{\Im}{\operatorname{Im}}
\fi

\ifdefined\arch
\renewcommand{\arch}{\operatorname{ar\,ch}}
\else
\newcommand{\arch}{\operatorname{ar\,ch}}
\fi

\ifdefined\arsh
\renewcommand{\arsh}{\operatorname{ar\,sh}}
\else
\newcommand{\arsh}{\operatorname{ar\,sh}}
\fi

\ifdefined\arth
\renewcommand{\arth}{\operatorname{ar\,th}}
\else
\newcommand{\arth}{\operatorname{ar\,th}}
\fi

\ifdefined\ch
\renewcommand{\ch}{\operatorname{ch}}
\else
\newcommand{\ch}{\operatorname{ch}}
\fi

\ifdefined\sh
\renewcommand{\sh}{\operatorname{sh}}
\else
\newcommand{\sh}{\operatorname{sh}}
\fi

\ifdefined\th
\renewcommand{\th}{\operatorname{th}}
\else
\newcommand{\th}{\operatorname{th}}
\fi

\ifdefined\Ln
\renewcommand{\Ln}{\operatorname{Ln}}
\else
\newcommand{\Ln}{\operatorname{Ln}}
\fi

\ifdefined\tg
\renewcommand{\tg}{\operatorname{tg}}
\else
\newcommand{\tg}{\operatorname{tg}}
\fi

\ifdefined\ctg
\renewcommand{\ctg}{\operatorname{ctg}}
\else
\newcommand{\ctg}{\operatorname{ctg}}
\fi

\ifdefined\intl
\renewcommand{\intl}{\int\limits}
\else
\newcommand{\intl}{\int\limits}
\fi

\ifdefined\ointl
\renewcommand{\ointl}{\oint\limits}
\else
\newcommand{\ointl}{\oint\limits}
\fi

\ifdefined\integrated
\renewcommand{\integrated}[3]{\left\{{#1}\right\}\left.\vphantom{#1}\right|_{#2}^{#3}}
\else
\newcommand{\integrated}[3]{\left\{{#1}\right\}\left.\vphantom{#1}\right|_{#2}^{#3}}
\fi

\ifdefined\pd
\renewcommand{\pd}[2]{\frac{\partial{#1}}{\partial{#2}}}
\else
\newcommand{\pd}[2]{\frac{\partial{#1}}{\partial{#2}}}
\fi

\ifdefined\rec
\renewcommand{\rec}[1]{\frac{1}{#1}}
\else
\newcommand{\rec}[1]{\frac{1}{#1}}
\fi

\ifdefined\gvec
\renewcommand{\gvec}[1]{\mbox{\boldmath${#1}$}}
\else
\newcommand{\gvec}[1]{\mbox{\boldmath${#1}$}}
\fi

\ifdefined\cvec
\renewcommand{\cvec}[1]{\mbox{\boldmath${#1}$}}
\else
\newcommand{\cvec}[1]{\mbox{\boldmath${#1}$}}
\fi

\ifdefined\td
\renewcommand{\td}[2]{\frac{d{#1}}{d{#2}}}
\else
\newcommand{\td}[2]{\frac{d{#1}}{d{#2}}}
\fi

\ifdefined\md
\renewcommand{\md}[2]{\frac{\mathrm{d}{#1}}{\mathrm{d}{#2}}}
\else
\newcommand{\md}[2]{\frac{\mathrm{d}{#1}}{\mathrm{d}{#2}}}
\fi

\ifdefined\z
\renewcommand{\z}[1]{\left({#1}\right)}
\else
\newcommand{\z}[1]{\left({#1}\right)}
\fi

\ifdefined\ae
\renewcommand{\ae}[1]{\left|{#1}\right|}
\else
\newcommand{\ae}[1]{\left|{#1}\right|}
\fi

\ifdefined\sz
\renewcommand{\sz}[1]{\left[{#1}\right]}
\else
\newcommand{\sz}[1]{\left[{#1}\right]}
\fi

\ifdefined\kz
\renewcommand{\kz}[1]{\left\{{#1}\right\}}
\else
\newcommand{\kz}[1]{\left\{{#1}\right\}}
\fi

\ifdefined\m
\renewcommand{\m}[1]{\mathrm{#1}}
\else
\newcommand{\m}[1]{\mathrm{#1}}
\fi

\ifdefined\c
\renewcommand{\c}[1]{\mathcal{#1}}
\else
\newcommand{\c}[1]{\mathcal{#1}}
\fi

\ifdefined\v
\renewcommand{\v}[1]{\mathbf{#1}}
\else
\newcommand{\v}[1]{\mathbf{#1}}
\fi

\ifdefined\Eq
\renewcommand{\Eq}[1]{Eq.~(\ref{#1})}
\else
\newcommand{\Eq}[1]{Eq.~(\ref{#1})}
\fi

\ifdefined\Eqs
\renewcommand{\Eqs}[2]{Eqs.~(\ref{#1}) and (\ref{#2})}
\else
\newcommand{\Eqs}[2]{Eqs.~(\ref{#1}) and (\ref{#2})}
\fi

\ifdefined\a
\renewcommand{\a}[1]{\aref({#1})}
\else
\newcommand{\a}[1]{\aref({#1})}
\fi

\ifdefined\A
\renewcommand{\A}[1]{\Aref({#1})}
\else
\newcommand{\A}[1]{\Aref({#1})}
\fi

\ifdefined\r
\let\R\r
\renewcommand{\r}[1]{(\ref{#1})}
\else
\newcommand{\r}[1]{(\ref{#1})}
\fi

\ifdefined\comm
\renewcommand{\comm}[2]{\left[{#1},{#2}\right]}
\else
\newcommand{\comm}[2]{\left[{#1},{#2}\right]}
\fi

\ifdefined\Follows
\renewcommand{\Follows}{\qquad\Rightarrow\qquad}
\else
\newcommand{\Follows}{\qquad\Rightarrow\qquad}
\fi

\ifdefined\follows
\renewcommand{\follows}{\quad\Rightarrow\quad}
\else
\newcommand{\follows}{\quad\Rightarrow\quad}
\fi

\ifdefined\followse
\renewcommand{\followse}{\quad\Rightarrow}
\else
\newcommand{\followse}{\quad\Rightarrow}
\fi

\ifdefined\bfollows
\renewcommand{\bfollows}{\Rightarrow\quad}
\else
\newcommand{\bfollows}{\Rightarrow\quad}
\fi

\ifdefined\equivalent
\renewcommand{\equivalent}{\quad\Leftrightarrow\quad}
\else
\newcommand{\equivalent}{\quad\Leftrightarrow\quad}
\fi

\ifdefined\obs
\renewcommand{\obs}[1]{\left\langle{#1}\right\rangle}
\else
\newcommand{\obs}[1]{\left\langle{#1}\right\rangle}
\fi

\ifdefined\ket
\renewcommand{\ket}[1]{\left|{#1}\right\rangle}
\else
\newcommand{\ket}[1]{\left|{#1}\right\rangle}
\fi

\ifdefined\bra
\renewcommand{\bra}[1]{\left\langle{#1}\right|}
\else
\newcommand{\bra}[1]{\left\langle{#1}\right|}
\fi

\ifdefined\braket
\renewcommand{\braket}[2]{\left<#1\vphantom{#2}\right|\left.#2\vphantom{#1}\right>}
\else
\newcommand{\braket}[2]{\left<#1\vphantom{#2}\right|\left.#2\vphantom{#1}\right>}
\fi

\ifdefined\ketbra
\renewcommand{\ketbra}[2]{\left|#1\vphantom{#2}\right>\left<#2\vphantom{#1}\right|}
\else
\newcommand{\ketbra}[2]{\left|#1\vphantom{#2}\right>\left<#2\vphantom{#1}\right|}
\fi

\ifdefined\matrixel
\renewcommand{\matrixel}[3]{\left<#1\vphantom{#2#3}\right|#2\left|#3\vphantom{#1#2}\right>} 
\else
\newcommand{\matrixel}[3]{\left<#1\vphantom{#2#3}\right|#2\left|#3\vphantom{#1#2}\right>} 
\fi

\ifdefined\contravcov
\renewcommand{\contravcov}[3]{{{#1}^{#2}_{}}_{#3}}
\else
\newcommand{\contravcov}[3]{{{#1}^{#2}_{}}_{#3}}
\fi

\ifdefined\covcontrav
\renewcommand{\covcontrav}[3]{{{#1}_{#2}^{}}^{#3}}
\else
\newcommand{\covcontrav}[3]{{{#1}_{#2}^{}}^{#3}}
\fi

\ifdefined\am
\renewcommand{\am}{{\hat{a}^{\vphantom\dagger}}}
\else
\newcommand{\am}{{\hat{a}^{\vphantom\dagger}}}
\fi

\ifdefined\ap
\renewcommand{\ap}{{\hat{a}^\dagger}}
\else
\newcommand{\ap}{{\hat{a}^\dagger}}
\fi

\ifdefined\bm
\renewcommand{\bm}{{\hat{b}^{\vphantom\dagger}}}
\else
\newcommand{\bm}{{\hat{b}^{\vphantom\dagger}}}
\fi

\ifdefined\bp
\renewcommand{\bp}{{\hat{b}^\dagger}}
\else
\newcommand{\bp}{{\hat{b}^\dagger}}
\fi

\ifdefined\arctg
\renewcommand{\arctg}{\mathrm{arc}\,\mathrm{tg}}
\else
\newcommand{\arctg}{\mathrm{arc}\,\mathrm{tg}}
\fi